\documentclass[preprint,showpacs,preprintnumbers,amsmath,amssymb]{revtex4-1}

\usepackage{amsmath, amssymb}
\usepackage[dvips]{graphicx}

\begin{document}
\title{Structure and modeling of the network of two-Chinese-character compound words in the Japanese language}
\author{Ken Yamamoto$^1$, Yoshihiro Yamazaki$^2$}
\affiliation{$^1$Department of Physics, Faculty of Science and Engineering, Chuo University, Kasuga, Bunkyo-ku, Tokyo 112-8851, Japan\\
$^2$Department of Physics, School of Advanced Science and Engineering, Waseda University, Okubo, Shinjuku-ku, Tokyo 169-8555, Japan}

\begin{abstract}
This paper proposes a numerical model
of the network of two-Chinese-character compound words
(two-character network, for short).
In this network, a Chinese character is a node 
and a two-Chinese-character compound word links two nodes.
The basic framework of the model is
that an important character gets many edges.
As the importance of a character,
we use the frequency of each character appearing in publications.
The direction of edge is given according to a random number assigned to nodes.
The network generated by the model is small-world and scale-free,
and reproduces statistical properties 
in the actual two-character network quantitatively.
\end{abstract}

\maketitle

\section{Introduction}
Research fields of network science are increasingly spreading,
and complex network analysis currently has become a fundamental piece
in understanding complex systems.
Network analysis in physics dates back to the discoveries of
the small-world \cite{Watts} and scale-free \cite{Barabasi} properties.
Small-world means the coexistence of small path length and high clustering,
and scale-free means
the power law of degree distribution: $P(k)\propto k^{-\gamma}$.
Some other features,
such as the community structure \cite{Girvan}, network motif \cite{Milo},
and hierarchical structure \cite{Ravasz},
have been developed.

Human languages are considered to be typical complex systems,
whose structures have been described by complex networks from various aspects:
thesaurus networks \cite{Motter}, word association networks \cite{Joyce}, word co-occurrence networks \cite{Dorogovtsev}, syntactic dependency networks \cite{Liu}, and so on.
Knowledge of language networks has been also applied to
text mining \cite{Matsuo},
natural language processing \cite{Mihalcea},
and language acquisition \cite{Sole}.

\begin{figure*}
\mbox{\raisebox{40mm}{(a)}}
\includegraphics[clip]{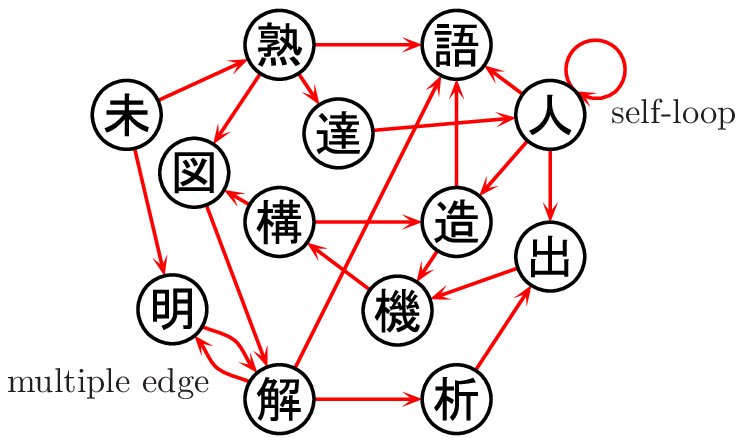}
\mbox{\raisebox{40mm}{(b)}}
\includegraphics[clip]{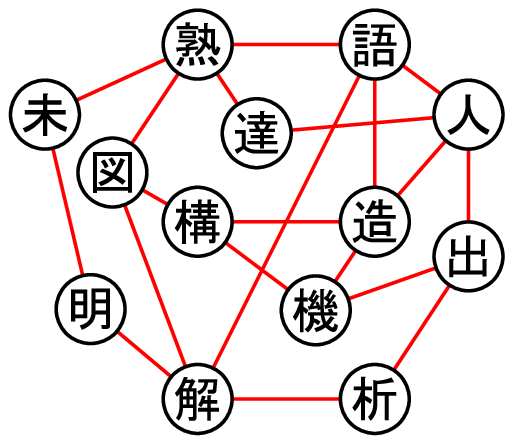}
\caption{
A small part of the two-character network of the \textit{Kojien} dictionary,
which is a directed network having multiple edges and self-loops (a).
The undirected counterpart (b) is also considered in this paper.
}
\label{fig1}
\end{figure*}

The present authors have studied the network structure of Chinese characters
in the Japanese language \cite{Yamamoto}.
The Japanese language has many words composed of two Chinese characters
(these words are called \textit{niji-jukugo} in Japanese).
For example, the characters ``\includegraphics{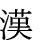}'' (Han or China) and ``\includegraphics{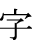}'' (character)
form the compound ``\includegraphics{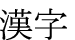}'' (Chinese character).
It was reported that the two-character words account for $70\%$
of the headwords of a Japanese-language dictionary \cite{Yokosawa}.
We constructed the \textit{two-character network}
by regarding each two-character compound 
as an edge connecting two Chinese characters (nodes).
Figure~\ref{fig1}(a) is a part of this network.
The first and second characters in a compound cannot be inverted generally,
so the network is directed;
the edge direction is indicated by an arrow 
from the first character to the second.
A few compounds can be invertible
[``\includegraphics{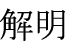}'' (clarify) and 
``\includegraphics{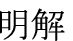}'' (clear and lucid) 
in Fig.~\ref{fig1}(a) for example],
and they form multiple edges and self-loops in a two-character network.
From the analysis of the undirected counterpart as in Fig.~\ref{fig1}(b),
we have previously found that 
networks built from headwords of the three Japanese-language dictionaries,
\textit{Kojien}, \textit{Iwanami Kokugo Jiten},
and \textit{Sanseido Kokugo Jiten} \cite{Dictionary},
are small-world and scale-free in common (Table~\ref{tbl1}).
A similar study was carried out in the Chinese language \cite{Peng}.
Small-world and scale-free properties are confirmed also in Chinese networks,
but their scale-free exponents 
$\gamma=1.40$ (Standard Chinese) and $1.49$ (Cantonese) are different from those of the Japanese networks.

The aim of this paper is to propose a stochastic model 
for reproducing statistical properties of the two-character network.
A basic point of the model is
that a Chinese character easily gets edges 
when it possesses high importance.
This model generates a directed network.
However, directed network analysis of two-character networks 
has not been done sufficiently in previous studies \cite{Yamamoto, Peng},
so we give directed network analyses of 
the HITS (hyperlink-induced topic search) algorithm 
and PageRank for a two-character network before the explanation of the model.
We confirm that the proposed model is quantitatively consistent with
the actual two-character network of \textit{Kojien}.

\begin{table*}\centering
\caption{
Fundamental characteristics of the three dictionaries.
The average path length $l$ is small,
and the clustering coefficient $C$ is much larger than $C_\mathrm{random}$
which is the average clustering coefficient of a random network having
the same number of nodes and edges as the actual network.
The power-law exponent $\gamma$ of the degree distribution
[$P(k)\propto k^{-\gamma}$] is also shown.
}
\label{tbl1}
\begin{tabular}{cccccccc}
\hline
Dictionary & Nodes & Edges & $\langle k\rangle$ & $l$ & $C$ & $C_\mathrm{random}$ & $\gamma$\\
\hline
\textit{Kojien} & 5458 & 74617 & 27.3 & 3.14 & 0.138 & 0.00501 & $1.14$\\
\textit{Iwanami} & 3904 & 32150 & 16.5 & 3.31 & 0.085 & 0.00424 & $1.16$\\
\textit{Sanseido} & 3444 & 28358 & 16.5 & 3.32 & 0.086 & 0.00483 & $1.14$\\
\hline
\end{tabular}
\end{table*}

\section{Directed network analysis: HITS and PageRank}\label{sec2}
In this section,
we apply the HITS algorithm and PageRank to the \textit{Kojien} network.
These two methods are originally developed to rate the importance of web pages 
based on hyperlinks,
and are now employed widely in analyses of many directed networks.

\begin{figure}\centering
\mbox{\raisebox{45mm}{(a)}}\includegraphics[clip]{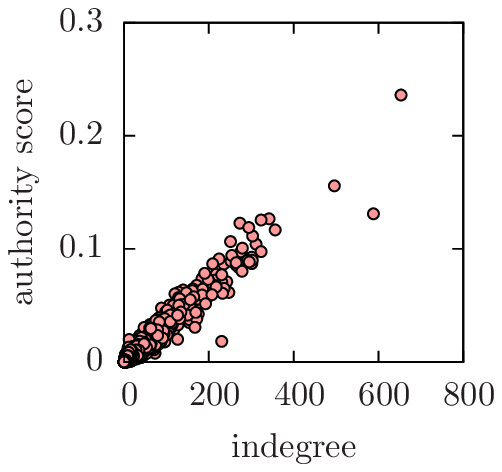}
\hspace{1cm}
\mbox{\raisebox{45mm}{(b)}}\includegraphics[clip]{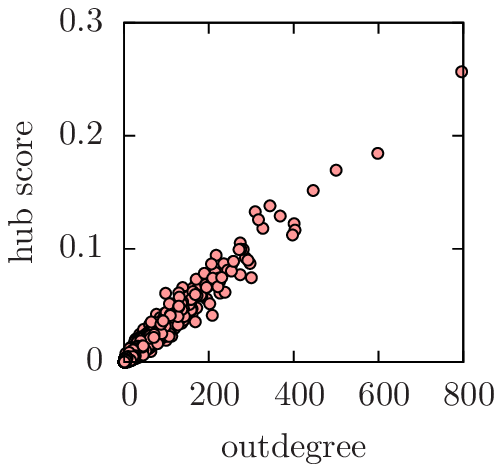}
\caption{
Results of the HITS algorithm.
Correlation between indegree and authority score (a), and
between outdegree and hub score (b).
These pairs are highly correlated.
}
\label{fig2}
\end{figure}

The HITS algorithm
assigns hub and authority scores to each node \cite{Kleinberg};
a node gets high hub score if it has many outgoing edges to nodes with high authority scores,
and a node gets high authority score if it has many incoming edges from nodes with high hub scores.
Figure~\ref{fig2} shows that
the authority score highly correlates with indegree [Panel (a)],
and that the hub score highly correlates with outdegree [Panel (b)].

Along with the HITS algorithm,
the PageRank is also a technique to assign the importance of each node in a directed network \cite{Brin}.
The idea of the PageRank is based on a relation that
a web page which receives many hyperlinks from many important pages
is a good page.
From the viewpoint of statistical physics,
the PageRank is a kind of the visiting probability of a random walk
which hops along the edge direction.
Technically, 
the random walker teleports with probability $q$ to an arbitrary node
so that the walker is not trapped in a node having no outgoing edges.
(The term ``teleport'' is officially used in researches of the PageRank.)
According to a study of the PageRank \cite{Chen},
people follow six hyperlinks on average until they begin new search,
and hence an appropriate teleport probability is $q\approx1/6$.
In fact, the standard value is $q=0.15$ 
in researches of web networks \cite{Brin, Chen, Langville}.
However, there is no clear advantage 
in considering a long sequence of nodes in our two-character network,
because one does not wander from character to character.
Only the structure of whether two nodes are linked or not is informative,
so we choose the teleport probability $q=1/2$ in this study.
The value $q=1/2$ has been used also in analysis of the citation network 
for the same reason \cite{Chen}.

\begin{figure}[t!]\centering
\mbox{\raisebox{45mm}{(a)}}\includegraphics[clip]{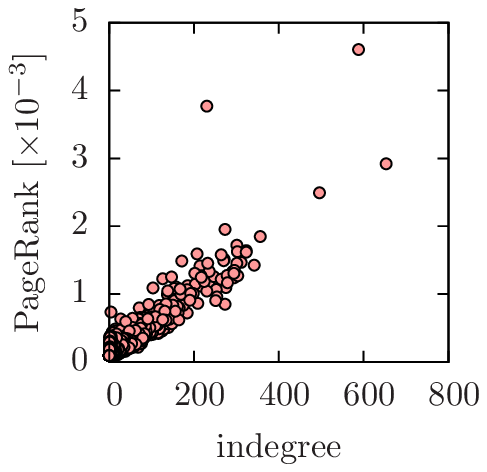}
\hspace{1cm}
\mbox{\raisebox{45mm}{(b)}}\includegraphics[clip]{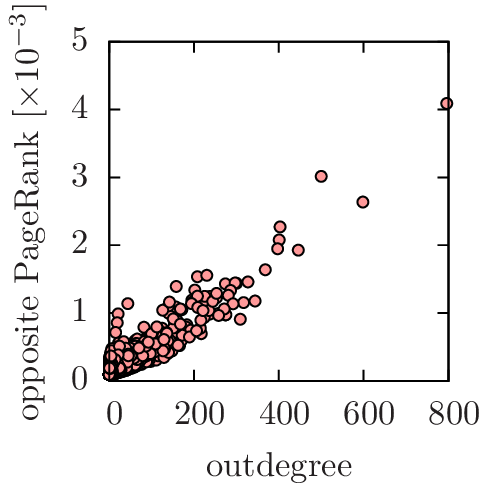}
\caption{
Correlation between indegree and the PageRank (a), and
between outdegree and the opposite PageRank (b).
}
\label{fig3}
\end{figure}

The PageRank measures the importance of a node based on how many good incoming edges are attracted.
For the two-character network, the importance of a node as the second character is measured by the PageRank.
As for the importance as the first character,
we calculate the ``opposite'' PageRank by inverting the directions of all the edges.
We show in Fig.~\ref{fig3} that the ordinary and opposite PageRanks correlate with indegree and outdegree, respectively.
The idea of the opposite PageRank is meaningless in a web network
because hyperlinks cannot be inverted,
but it has been applied to some other networks.
For instance, in the study of an inter-firm network,
the ordinary network represents the flow of money,
and the opposite network represents the flow of material and service \cite{Ohnishi}.

\begin{table}[t!]\centering
\caption{
Correlation coefficients of in- and out-degrees
versus authority score, hub score, PageRank, and opposite PageRank.
}
\begin{tabular}{r|r|r||r|r|}
& authority & hub & PageRank & \shortstack{opposite\\ PageRank}\\
\hline
indegree & 0.972 & 0.589 & 0.917 & 0.425\\
\hline
outdegree & 0.580 & 0.980 & 0.484 & 0.905\\
\hline
\end{tabular}
\label{tbl2}
\end{table}

The correlation coefficients are summarized in Table~\ref{tbl2}.
Correlation coefficients of indegree-authority, indegree-PageRank,
outdegree-hub, and outdegree-opposite PageRank are nearly 1,
and the others are also positive but relatively low.

\section{Modeling of the two-character network}\label{sec3}
We propose a numerical model for the two-character networks here.
The model generates a directed network,
and we discuss not only undirected properties shown in Table~\ref{tbl1}
but also directed ones stated in the previous section.
We introduce the following two properties into the model.

The first point is that
importance or utility of each character is not uniform;
there are a few very general and important characters
together with a great number of characters used in only specific situations.
We assume that the importance of a Chinese character $i$
is represented by a positive number $x_i$.
Intuitively,
two characters $i$ and $j$ easily form a compound
when the product $x_ix_j$ is large.
For characterizing the importance of each character,
we employed a survey on the frequency 
that each Chinese character appears in Japanese publications,
conducted by the Agency for Cultural Affairs of Japan \cite{Bunkacho}.
The survey enumerates 8576 different characters,
in 49072315 Chinese characters appearing in 860 publications (books, magazines, and textbooks).
The average frequency is 5722,
and the character ``\includegraphics{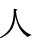}'' (person) has the largest frequency 610660.
It is natural to consider that a character of large frequency
has strong ability in formation of compound words.
Therefore, we regard the frequency of each character
as the importance $x_i$.

\begin{figure}[t]\centering
\mbox{\raisebox{50mm}{(a)}}
\includegraphics[clip]{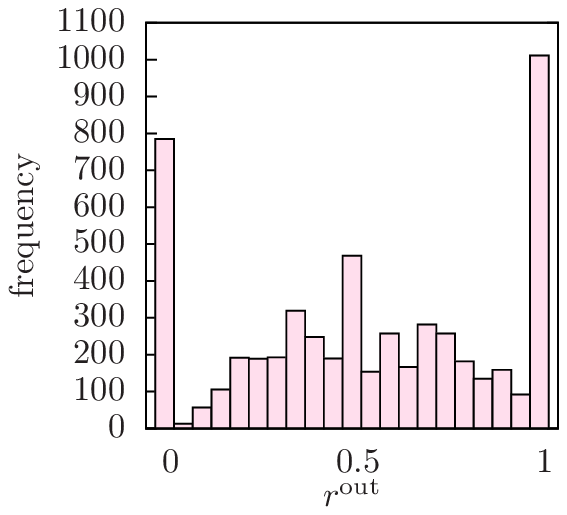}
\hspace{1cm}
\mbox{\raisebox{50mm}{(b)}}
\includegraphics[clip]{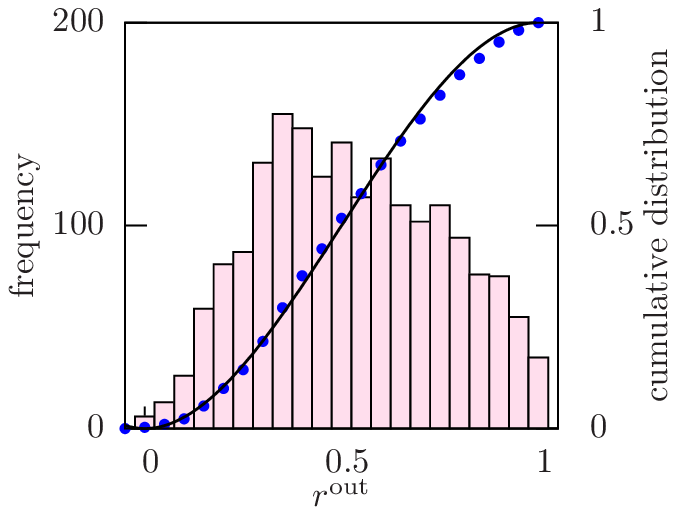}
\caption{
(a) The histogram of the ratio $r^\mathrm{out}$ of the outdegree to the total degree in the directed network of \textit{Kojien}.
(b) The histogram of $r^\mathrm{out}$ by the nodes 
whose degrees are more than 20.
The cumulative distribution (points) is approximated well 
by the parabolic distribution (curve).
}
\label{fig4}
\end{figure}

The second point is the edge direction.
We define $r^\mathrm{out}_i$ as the ratio of the outdegree to the total degree
of node $i$,
i.e., $r^\mathrm{out}_i=k^\mathrm{out}_i/k_i$.
The ratio $r^\mathrm{out}_i$ represents strength that the node $i$ becomes the first character in a compound.
From the viewpoint of linguistics \cite{Nomura, Todo},
the character position (first or second) in a compound is closely related to 
formation principle and structure of compound words;
for example, the characters ``\includegraphics{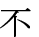}'' and ``\includegraphics{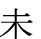}'' used for negation
usually become the first and hence their $r^\mathrm{out}$'s are close to 1.
The frequency histogram of $r^\mathrm{out}$ of the directed \textit{Kojien} network is illustrated in Fig.~\ref{fig4}(a).
It seems
that there are many nodes of $r^\mathrm{out}=0$ and $r^\mathrm{out}=1$,
but these peaks are mainly created by nodes of degree 1, which can take only $r^\mathrm{out}=0$ or 1.
Similarly, the peak at $r^\mathrm{out}=0.5$ is mainly due to a large number of nodes of degree 2.
In order to reduce statistical effects by low-degree nodes,
we show in Fig.~\ref{fig4}(b) a histogram of $r^\mathrm{out}$ 
only from the nodes whose degrees are more than 20.
We have heuristically found that the \textit{parabolic} probability distribution between 0 and 1 is appropriate for $r^\mathrm{out}$.
[The probability density of the parabolic distribution is $f_\mathrm{parabolic}(r)=6r(1-r)$,
and the lower cumulative distribution is $F_\mathrm{parabolic}(r)=3r^2-2r^3$.]
The parabolic distribution is better than uniform, Gaussian, and triangular distributions, but so far, we have no idea why it works well.

We propose a model based on these two properties.
We set 8576 nodes,
where the node $i$ is assigned the importance $x_i$
and the strength $r^\mathrm{out}_i$ to become the first character;
$x_i$ is the frequency recorded in the survey \cite{Bunkacho},
and $r^\mathrm{out}_i$ is a random number drawn from the parabolic distribution on the unit interval $[0,1]$.
We assume that $x_i$ and $r^\mathrm{out}_i$ have no correlation.

The model comprises two steps
to determine whether two nodes $i$ and $j$ are connected or not.
The first step is involved with importance $x_i$ and $x_j$.
With probability $p_{ij}=cx_ix_j$, the pair $(i, j)$ passes the first step,
and proceeds to the second step.
The constant $c$ is determined later.
Otherwise, no edge is created between $i$ and $j$.
If $p_{ij}$ is greater than 1, the pair $(i, j)$ unconditionally proceeds to the second step.
The link probability $p_{ij}$ is similar to
that considered in the fitness model \cite{Caldarelli},
and we comment on the relevance of our model to the fitness model 
in Section~\ref{sec5}.

\begin{figure}[t!]\centering
\includegraphics[clip]{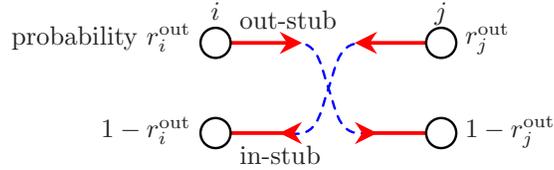}
\caption{
Illustration of the second step of the model.
The node $i$ puts forth the out- or in-stub with probability
$r^\mathrm{out}_i$ and $1-r^\mathrm{out}_i$, respectively.
The node $j$ also puts forth a stub.
The two nodes are linked when the stubs have opposite directions
(indicated by dashed curves).
}
\label{fig5}
\end{figure}

In the second step,
each of the two nodes $i$ and $j$ sprouts
an incomplete half edge (called a \textit{stub} in network analysis).
The node $i$ puts forth either 
an out-stub with probability $r^\mathrm{out}_i$ or
an in-stub with $1-r^\mathrm{out}_i$;
the node $j$ also puts an out- or in-stab with $r^\mathrm{out}_j$ and
$1-r^\mathrm{out}_j$.
The two nodes $i$ and $j$ become linked if the two stubs have opposite directions, and otherwise the stubs are dead.
Therefore, in this second step,
an edge from $i$ to $j$ is created
with probability $r^\mathrm{out}_i(1-r^\mathrm{out}_j)$,
an edge from $j$ to $i$ is created 
with $(1-r^\mathrm{out}_i)r^\mathrm{out}_j$,
and no edge between $i$ and $j$ is created otherwise.
We illustrate this step in Fig.~\ref{fig5}.

Until all possible pairs of nodes $i$ and $j$ ($1\le i<j\le8576$) are checked,
we determine one by one whether an edge between $i$ and $j$ is created or not 
by applying the two steps.
A node having very small importance cannot get edges usually.
We exclude such a node from the final model-generated network.
More precisely,
we focus on the largest connected component in this model.
In principle, multiple edges and self-loops cannot be created in this model,
but this is not serious because they are few in actual networks.
The calculation results of this model are presented in the next section.

\section{Calculation results of the model}\label{sec4}
\begin{figure}[t!]\centering
\includegraphics[clip]{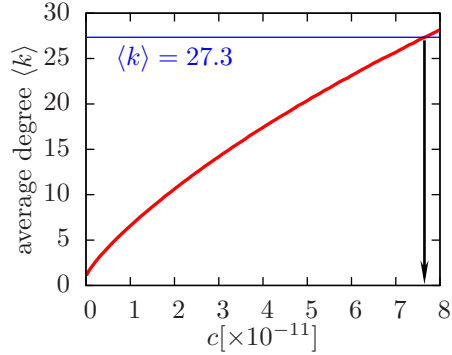}
\caption{
Determination of the coefficient $c$ of the link probability $p_{ij}$.
We choose $c=7.65\times10^{-11}$ so that 
the average degree of the model network 
becomes close to $\langle k\rangle=27.3$ of the \textit{Kojien} network.
}
\label{fig6}
\end{figure}

Initially, we need to determine the coefficient $c$ of the link probability.
When $c$ becomes larger, nodes get more edges to make the network larger and denser.
We choose the value of $c$
so that the average degree of the model network becomes close to
that of the \textit{Kojien} network ($\langle k\rangle=27.3$).
We adjust the average degree because it is a fundamental quantity for network structure;
for instance, the Erd\H{o}s-R\'enyi random graph goes through
a phase transition at the critical average degree \cite{Durrett}.
Figure~\ref{fig6} is the numerical result 
of the average degree $\langle k\rangle$ 
of the model network as a function of $c$,
and the appropriate value is $c=7.65\times10^{-11}$.

\begin{figure}[t!]\centering
\mbox{\raisebox{45mm}{(a)}}\includegraphics[clip]{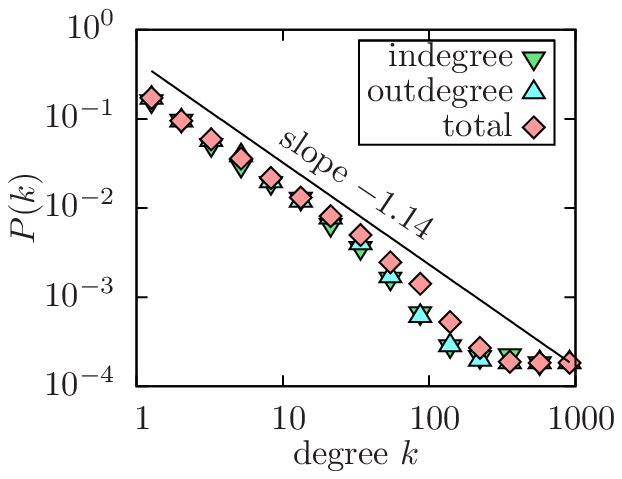}
\mbox{\raisebox{45mm}{(b)}}\includegraphics[clip]{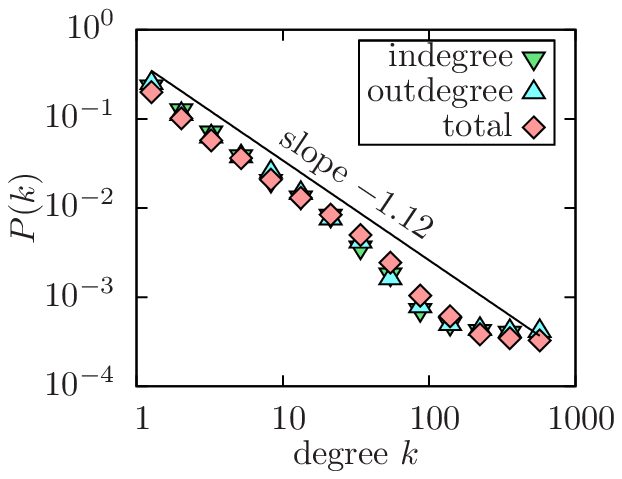}
\caption{
Directed degree distributions of the \textit{Kojien} network (a)
and one numerical sample of the model network (b).
Indegree, outdegree, and total degree (sum of in- and out-degrees)
are plotted.
}
\label{fig7}
\end{figure}

Table~\ref{tbl3} shows undirected properties of the model network
and the \textit{Kojien} network.
The row of ``Model'' is a numerical result averaged over 1000 samples.
The average path length $l$, clustering coefficient $C$, and power-law exponent $\gamma$ obtained by the model are very close to those of \textit{Kojien}.
We made 8576 nodes initially,
but on average $5514(=8576-3062)$ nodes have no chance to get edges.
In the row titled ``Model (uniform $r^\mathrm{out}$) of Table~\ref{tbl3}, 
the network properties with $r^\mathrm{out}$ distributed uniformly on $[0,1]$ is also presented.
We claim that
the network structure does not depend largely on the distribution of $r^\mathrm{out}$.
We show degree distributions of the \textit{Kojien} network and the model network in Fig.~\ref{fig7}.

The parameter $c$ is used for the adjustment of the average degree,
and there are no control parameters left in the model.
Consequently, it is natural that the numbers of nodes and edges of the model
differ from those of the \textit{Kojien} network;
we should regard the model-generated network as a miniature of \textit{Kojien}.
It is rather surprising that the model produces $l$, $C$, and $\gamma$ quite well.
This good result implies that the proposed model captures essential features
of the actual two-character network.
The agreement of the exponent $\gamma$ has a further meaning
related to the fitness model.
See Section~\ref{sec5} for discussion of this.

\begin{table}[t!]\centering
\caption{
Comparison of network characteristics between the model and \textit{Kojien},
where $\langle k\rangle=27.3$.
The values of $l$, $C$, and $\gamma$ by the model
are in good agreement with those of the \textit{Kojien} network.
The row titled ``Model (uniform $r^\mathrm{out}$)'' is a numerical result 
with which $r^\mathrm{out}$ is distributed uniformly on the interval $[0,1]$.
}
\label{tbl3}
\begin{tabular}{lcccccc}
\hline
 & Nodes & Edges & $l$ & $C$ & $\gamma$\\
\hline
\textit{Kojien} & 5458 & 74617 & 3.14 & 0.138 & $1.14$\\
Model & 3062 & 41841 & 2.86 & 0.145 & $1.12$\\
Model (uniform $r^\mathrm{out}$) & 3060 & 41826 & 2.86 & $0.141$ & 1.12\\
\hline
\end{tabular}
\end{table}

\begin{figure*}\centering
\mbox{\raisebox{45mm}{(a)}}\includegraphics[clip]{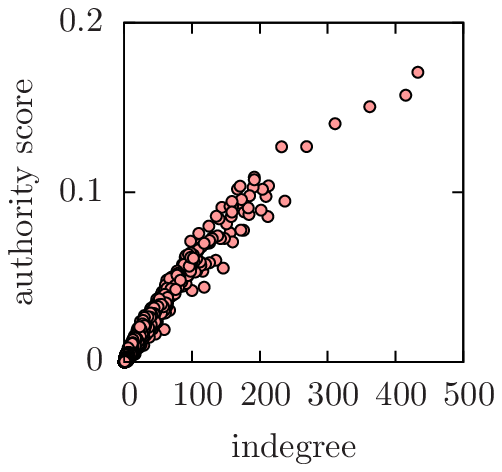}
\hspace{1cm}
\mbox{\raisebox{45mm}{(b)}}\includegraphics[clip]{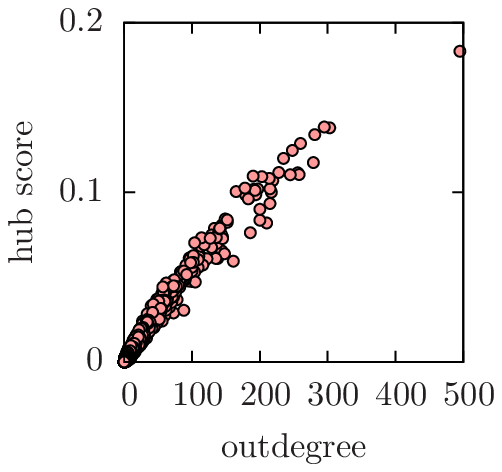}
\\
\mbox{\raisebox{45mm}{(c)}}\includegraphics[clip]{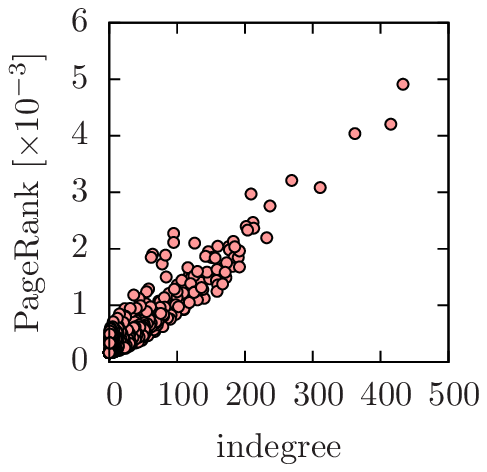}
\hspace{1cm}
\mbox{\raisebox{45mm}{(d)}}\includegraphics[clip]{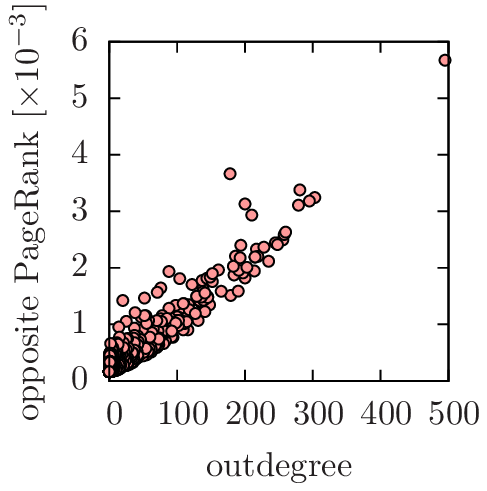}
\caption{
Results of directed network analyses of one sample network.
Indegree correlates with the authority score (a) and the PageRank (c),
and outdegree correlates with the hub score (b) and the opposite PageRank (d).
}
\label{fig8}
\end{figure*}

Next we compare directed properties.
As shown in Fig.~\ref{fig8},
indegree correlates with the authority and PageRank scores,
and outdegree with the hub and opposite PageRank scores.
The correlation coefficients of the model (Table~\ref{tbl4})
have the same tendency as those of the \textit{Kojien} network (Table~\ref{tbl2}).

Judging from the numerical results comprehensively,
we conclude that the proposed model successfully describes the two-character network.

\begin{table}[t!]\centering
\caption{
Correlation coefficients of one sample network of the model.
Indegree strongly correlates with the authority and PageRank scores,
and outdegree correlates with hub the and opposite PageRank scores.
The other correlations are low.
}
\begin{tabular}{r|r|r||r|r|}
& authority & hub & PageRank & \shortstack{opposite\\ PageRank}\\
\hline
indegree & 0.979 & 0.381 & 0.928 & 0.326\\
\hline
outdegree & 0.369 & 0.983 & 0.343 & 0.935\\
\hline
\end{tabular}
\label{tbl4}
\end{table}

\section{Discussion}\label{sec5}
\begin{figure}[b!]\centering
\includegraphics[clip]{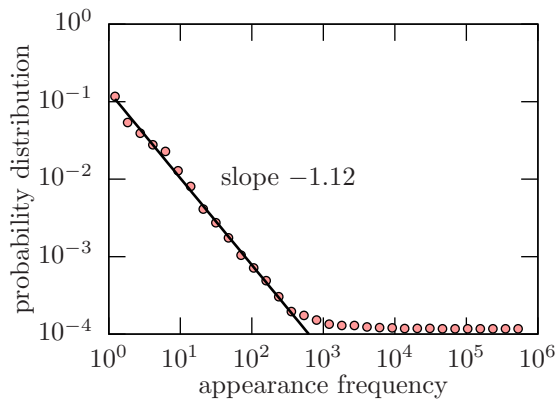}
\caption{
The probability distribution of the frequency in the survey \cite{Bunkacho}.
A power law with exponent $-1.12$ holds for frequencies less than $10^3$.
}
\label{fig9}
\end{figure}

We give further consideration to the scale-free property of our model
by making a comparison with the fitness model.
In the fitness model \cite{Caldarelli},
the node $i$ is assigned a positive value $x_i$ called the fitness,
and two nodes $i$ and $j$ are linked with probability proportional to the product $x_ix_j$.
If the fitness of a node is drawn from the power-law distribution
$\rho(x)\propto x^{-\beta}$,
the generated network is known to become scale-free having degree distribution
$P(k)\propto k^{-\beta}$.
In order to show that our model is similar to the fitness model,
we check the distribution of frequency of each Chinese character listed in the survey \cite{Bunkacho}.
As in Fig.~\ref{fig9},
the distribution of frequency follows a power-law distribution 
with exponent $-1.12$, which is the same value as $\gamma$ in Table~\ref{tbl3}.
Therefore, we conclude that our model is very close to the fitness model;
a major difference is that the fitness model usually does not take the edge direction into account.
We believe that a \textit{directed} fitness model is useful not only in linguistic systems.

The two-character network can be regarded as a kind of co-occurrence network,
in the sense that two Chinese characters are linked when they appear in the same compound word.
However, the scale-free exponents are different in the two-character network and the Chinese character co-occurrence network;
in the Chinese language,
two-character networks have $\gamma=1.40$ and $1.49$ \cite{Peng}, and co-occurrence networks have $\gamma$ between $1.98$ and $2.35$ \cite{Liang}.
In the co-occurrence network, two characters are linked 
when they appear successively in a sentence;
hence, one sentence can form more than one edges.
Yet in the two-character network, one compound word forms just one edge.
We think that this constraint on the compound words causes the difference of exponents.

Studies on the syllable networks 
in Portuguese \cite{Soares} and Chinese \cite{Peng} inferred 
that the power-law degree distributions of these networks are attained by
preferential attachment, that is,
a node with high degree tends to attract many edges afterward.
The preferential-attachment mechanism involves the growth of a network,
and the verification of this hypothesis requires 
the evolution of word creation.
As far as the authors know, there seems to be no complete and useful data source about when each word was created,
so we think that preferential attachment in language evolution 
is difficult to verify even in a qualitative level.
Furthermore,
an extended model based on preferential attachment 
can produces networks with $2<\gamma<\infty$ \cite{DMS}.
The actual scale-free exponents $\gamma=1.40$ and $1.49$ of Chinese networks \cite{Peng} are out of this range.
Hence, preferential attachment seems not to be plausible in two-character networks.
Our model, by contrast, makes a network similar to the actual one
without considering the evolution of words.
The key is the characterization of the character importance;
we used official data on the frequency of the character appearing in Japanese publications.
Our model is static, and assigns the character importance in advance.
This is different from the preferential-attachment model,
where the importance of a node is given by when the node is added to the network.
In terms of linguistics,
our static model is comparable with a synchronic study,
and a preferential-attachment model with a diachronic study.

We believe that our model is applied to languages other than Japanese
to understand their network structure.
For example, the difference of the scale-free exponent
in Japanese and Chinese two-character networks will be explained from
the difference of the power-law exponent in the character-importance distribution.

\section{Conclusion}
Measurement and modeling of a network of two-character compound in Japanese 
are presented in this paper.
In Section \ref{sec2}, we have measured directed properties of an actual two-character network, which was previously reported to be small-world and scale-free.
With the HITS and PageRank algorithms,
we have found that indegree highly correlates with authority and PageRank scores, and that outdegree with hub and opposite PageRank scores.
In Section \ref{sec3}, we have proposed a numerical model for reproducing properties of the two-character network.
We have adopted a basic formation principle of a two-character word that an important character is easy to get many edges;
we have used an official data set of the appearance frequency in publications for the importance of a character.
For the edge direction in the model, we have introduced the strength $r^\mathrm{out}$ to become the first character in a two-character word.
We note that the data set of character frequency in publications is independent from the network characteristics,
and that a choice of the distribution of $r^\mathrm{out}$ does not have a large effect on network statistics.
Therefore, in the formulation and calculation of the model,
we have effectively used only the average degree $\langle k\rangle$ for determining the parameter $c$ in the model,
and have not used the other network characteristics.
Nevertheless, we have successfully made a comparison between the actual network of \textit{Kojien} and the model-generated network, as discussed in Section \ref{sec4}.
We conclude that our modeling is simple but to the point.

\acknowledgments
The authors are very grateful to Mr.~Yasuhiro Takeda
and the Agency for Cultural Affairs of Japan
for replying a query about a survey and offering a copy kindly.

\end{document}